\documentclass[twocolumn]{aastex62}
\usepackage{graphicx}
\usepackage{color}
\usepackage{amsmath}
\usepackage{lineno}
\usepackage{hyperref}

\hypersetup{
    pdfnewwindow=true,      
    colorlinks=true,        
    linkcolor=blue,         
    citecolor=blue,        
    filecolor=blue,         
    urlcolor=blue           
}

\providecommand{\aap}[0]{Astron. Astrophys. }

\providecommand{\apjl}[0]{Astrophys. J. Lett. }
\providecommand{\apjs}[0]{Astrophys. J. Supp. Ser. }

\definecolor{carminered}{rgb}{1.0, 0.0, 0.22}
\newcommand*{\Szabi}{\textcolor{black}}

\definecolor{amethyst}{rgb}{0.6, 0.4, 0.8}

\begin{document}

\title{Early Solar System $r$-process Abundances Limit Collapsar Origin}

\author{I. Bartos}
\thanks{imrebartos@ufl.edu}
\affiliation{Department of Physics, University of Florida, PO Box 118440, Gainesville, FL 32611-8440, USA}
\author{S. M\'arka}
\affiliation{Department of Physics, Columbia University, 550 W 120th St., New York, NY 10027, USA}

\begin{abstract}
Heavy elements produced exclusively through rapid neutron capture (the '$r$-process') originate from violent cosmic explosions. While neutron star mergers are the primary candidates, another plausible production site are 'collapsars'---collapsing massive stars that form a black hole with an accretion disk. Here we show that collapsars are too rare to be the prime origin of $r$-process elements in the Solar System. By comparing numerical simulations with the early Solar System abundances of actinides produced exclusively through the $r$-process, we exclude higher than 20\% contribution from collapsars with 90\% confidence. We additionally limit $r$-process ejecta masses from collapsars to less than 10\% of the ejecta mass from neutron star mergers, about $10^{-2}$\,M$_\odot$. 
\end{abstract}


\section{Introduction} 

The cosmic creation of heavy $r$-process elements is still not well understood. Core-collapse supernovae, which were historically considered to be the main source \citep{1957RvMP...29..547B,1994ApJ...433..229W}, are too frequent to explain the measured isotopic abundances in the early Solar System, in deep-sea sediments and in metal-poor stars \citep{2019Natur.569...85B,2015NatCo...6E5956W,2015NatPh..11.1042H,2018ApJ...860...89M}. 

Neutron star mergers are natural candidates as a major production site. They eject high-density, neutron-rich matter in sufficient quantities to be the main source of Galactic $r$-process elements \citep{2011ApJ...738L..32G,2015ApJ...807..115S,2015MNRAS.447..140V}. This possibility is further corroborated by recent multi-messenger observations of a neutron star merger accompanied by a kilonova \citep{2017ApJ...848L..12A,2017Sci...358.1556C,2017ApJ...848L..12A}. The rate of neutron star mergers in the Milky Way is $10-100$\,Myr$^{-1}$ \citep{2019ApJ...870...71P}, about a 1000 times less than the rate of supernovae \citep{1994ApJS...92..487T}, which is consistent with expectations from isotopic abundances in the early Solar System and deep-sea sediments \citep{2015NatCo...6E5956W,2015NatPh..11.1042H,2019Natur.569...85B}.   

Rare source types other than neutron star mergers may also contribute to $r$-process enrichment. Collapsars are stellar core-collapse events that have sufficiently massive cores to form black holes, and sufficient angular momentum at the time of collapse to form an accretion disk \citep{1999ApJ...524..262M}. The resulting accreting black holes could produce $r$-process elements through disk winds similarly to neutron star mergers \citep{2004ApJ...606.1006P,2005ApJ...629..341K,MetzgerCollapsar}. 

Collapsars are only expected in stars with low metallicities, while neutron star mergers are delayed compared to star formation. Therefore, a collapsar origin could help explain the presence of $r$-process elements in extremely metal poor stars, which is more difficult with neutron star mergers \citep{2016Natur.531..610J,2017ApJ...836..230C,2019ApJ...876...28S}. On the other hand, the observed $r$-process enrichment of some metal-poor stars disfavors sources like collapsars that are significant metal producers \citep{2019ApJ...877L..24M}. Additional challenges to the neutron star mergers as the main $r$-process production site include the strong enhancements in heavy $r$-process elements of some stars in ultra faint dwarf galaxies \citep{2016Natur.531..610J,2017ApJ...838...44H}, as the low escape velocities and short star-formation epochs in such galaxies would require unusually small natal kick velocities and fast merger time \citep{2019arXiv190109044S}. Further, Galactic $r$-process enrichment at high metallicities appears to be slower than expected from neutron star mergers, although this is dependent on the uncertain event rate and time delay compared to star formation \citep{2017ApJ...836..230C,2018IJMPD..2742005H}.

We examined the origin of $r$-process elements in the early Solar System by considering fractional contributions from both neutron star mergers and collapsars. We used the abundances of short-lived radioactive isotopes that encode information on their production and deposition history. Even though these elements are by now extinct in the Solar System, meteorites that condensed during the early Solar System still carry their imprint \citep{2006mess.book..127N}. The early Solar System abundance of a short-lived radioactive isotope ($N_{\rm SLR}$) can be estimated by comparing the abundance of its decay product with that of a chemically identical isotope ($N_{\rm stable}$). The measured abundance ratio $N_{\rm SLR}/N_{\rm stable}$ relative to the elements' production ratio $P_{\rm SLR}/P_{\rm stable}$ tells us the time interval between the astrophysical event that synthesized the elements and the formation of the Solar System.

\section{Methods} 

\subsection{Abundances} 

The ratios of the abundances of different elements produced in the ejecta are taken to be their production ratios. Here, we adopted estimates of production ratios found in \cite{2016SciA....2E1400T} (see also \citealt{2019Natur.569...85B}). For $^{247}$Cm and $^{244}$Pu we used another actinide, $^{232}$Th, as our reference isotope. $^{232}$Th is long-lived with $t_{1/2}=1.4\times10^{10}$yr, therefore it is still detectable in remnants from the Early Solar System. Our reference isotope for $^{129}$I is another $r$-process product, $^{127}$I, which is stable. For $^{129}$I for which \cite{2016SciA....2E1400T} 
obtained the ratios by averaging the results of multiple previous studies, we adopted the mean difference between the average value and the values of these studies. 

We adopted early Solar System abundance ratios for $^{247}$Cm and $^{244}$Pu from \cite{2016SciA....2E1400T}, and for $^{129}$I from \cite{LUGARO20181}.

\subsection{Collapsar rate in the Milky Way} 

Long GRBs occur in low-metallicity, highly star-forming environments \citep{2006ApJ...637..914W}. Ongoing star formation is important since massive stars that produce collapsars only live for a few million years, while low-metallicity limits stellar winds that would otherwise reduce the star's mass before collapse. These requirements make collapsars rare in the Milky Way \citep{2006Natur.441..463F,2006ApJ...638L..63L}. 

To estimate the Galactic rate of collapsars as a function of time, we consider its dependence on metallicity and star formation rate in the Milky Way. For fixed metallicity, we assume that the collapsar rate is proportional to the core-collapse supernova rate. Based on the fraction of stars in the Milky Way that have low metallicities similar to long-GRB host galaxies, the Galactic collapsar rate at present is expected to be about 5\% of the collapsar rate in the local universe implied by star formation only \citep{2014PhRvL.113w1102P}. We extend this 95\% metallicity-suppression by assuming that the collapsar rate is suppressed by $0.95Z(t)/Z(t_{0})$, where $Z(t)$ is the Galactic metallicity at time $t$, with $t_0$ being the present day. The precise shape of this suppression does not meaningfully affect our results. We adopt the metallicity evolution of the Milky Way from \cite{2015ApJ...808..132H} (see their Fig. 9. We chose their result at 4\,kpc as that region has the highest star formation; the difference is not large at different radii).

We assume that all collapsars produce a long gamma-ray burst. We take a local long gamma-ray burst rate of 1.3\,Gpc$^{-3}$yr$^{-1}$ \citep{2010MNRAS.406.1944W}. Using a characteristic long gamma-ray burst beaming factor of $5\times10^{-3}$ defined as the fraction of the sky in which a gamma-ray burst can be observed from cosmological distances \citep{2016ApJ...818...18G}, the corresponding local collapsar rate is $\sim260$ \,Gpc$^{-3}$yr$^{-1}$. The local star formation rate density is $\sim 10^{7}$\,M$_\odot$Gpc$^{-3}$yr$^{-1}$ \citep{2012A&A...539A..31C}, while the present star formation rate in the Milky Way is $\sim1$\,M$_{\odot}$yr$^{-1}$ \citep{1998ApJ...507..229P}. 

Combining these, we arrive at a present Galactic collapsar rate of $\sim1$\,Myr$^{-1}$. Beyond taking into account the metallicity-suppression above, we computed past collapsar rate by additionally taking into account the evolution of the core-collapse supernova rate in the Milky Way. We adopted the Galactic core collapse supernova rate as a function of time based on the high resolution, zoom-in cosmological simulation of a Milky Way Galaxy analog, called Eris \citep{2011ApJ...742...76G}, computed by \cite{2015ApJ...807..115S}.

\subsection{Neutron star merger rate in the Milky Way} 

We adopted the Galactic rate of neutron star mergers as a function of time based on the Eris simulation \citep{2011ApJ...742...76G}, computed by \cite{2015ApJ...807..115S}.  This rate varies within about $5-10$\,Myr$^{-1}$ over the history of the Milky Way, consistent with population synthesis estimates \citep{2019arXiv190300083A,2015ApJ...807..115S}.

\subsection{Monte Carlo simulations} 

In one Monte Carlo realization of the Milky Way, neutron star mergers and collapsars were randomly placed in space and time using the above rates throughout the lifetime of the Galaxy. The spatial probability distribution of neutron star merger followed the Galactic stellar mass distribution \citep{2011MNRAS.414.2446M}. Collapsars were placed randomly following a probability density radially proportional to the Galactic star formation rate, with no azumithal dependence, and with a scale height of 80\,pc \citep{2010ApJ...710L..11R}. 

We simulated the chemical mixing of $r$-process elements following \cite{2015NatPh..11.1042H}, accounting for radioactive decay and turbulent diffusion within the Milky Way with diffusion coefficient
\begin{equation}
D\approx0.1\, \mbox{kpc}^{2}\,\mbox{Gyr}^{-1}\,\left(\frac{\alpha}{0.1}\right)\left(\frac{v_{\rm t}}{7\,\mbox{km\,s}^{-1}}\right)\left(\frac{H}{0.2\,\mbox{kpc}}\right)
\end{equation}
where $\alpha$ is the mixing length parameter, $v_{\rm t}$ is the typical turbulence velocity in the interstellar medium, and $H$ is the interstellar-medium scale height. We adopt $D=0.1$\,kpc$^2$Gyr$^{-1}$ below following \cite{2015NatPh..11.1042H} (see also \citealt{2012ApJ...758...48Y}). We used identical ejecta mass and composition for all mergers and, independently, all collapsars. Abundance ratios in the early Solar System were taken to be proportional to abundance ratios in the interstellar medium near the pre-Solar nebula at the time of the formation of the Solar System. We assumed that the time between deposition into the pre-Solar nebula and the condensation of meteorites that preserved the imprint of short-lived isotopes in the early Solar System is negligible \citep{2017GeCoA.207....1T,2011AREPS..39..351D}.

We obtained the probability distribution of actinide abundance ratios in the early Solar System by varying the time when the Solar System was formed between $8-9$\,Gyr within a Monte Carlo realization of the Milky Way, and by computing $10^3$ realizations. 

\section{Results} 

\begin{figure}
\centering
\includegraphics[width=0.49\textwidth]{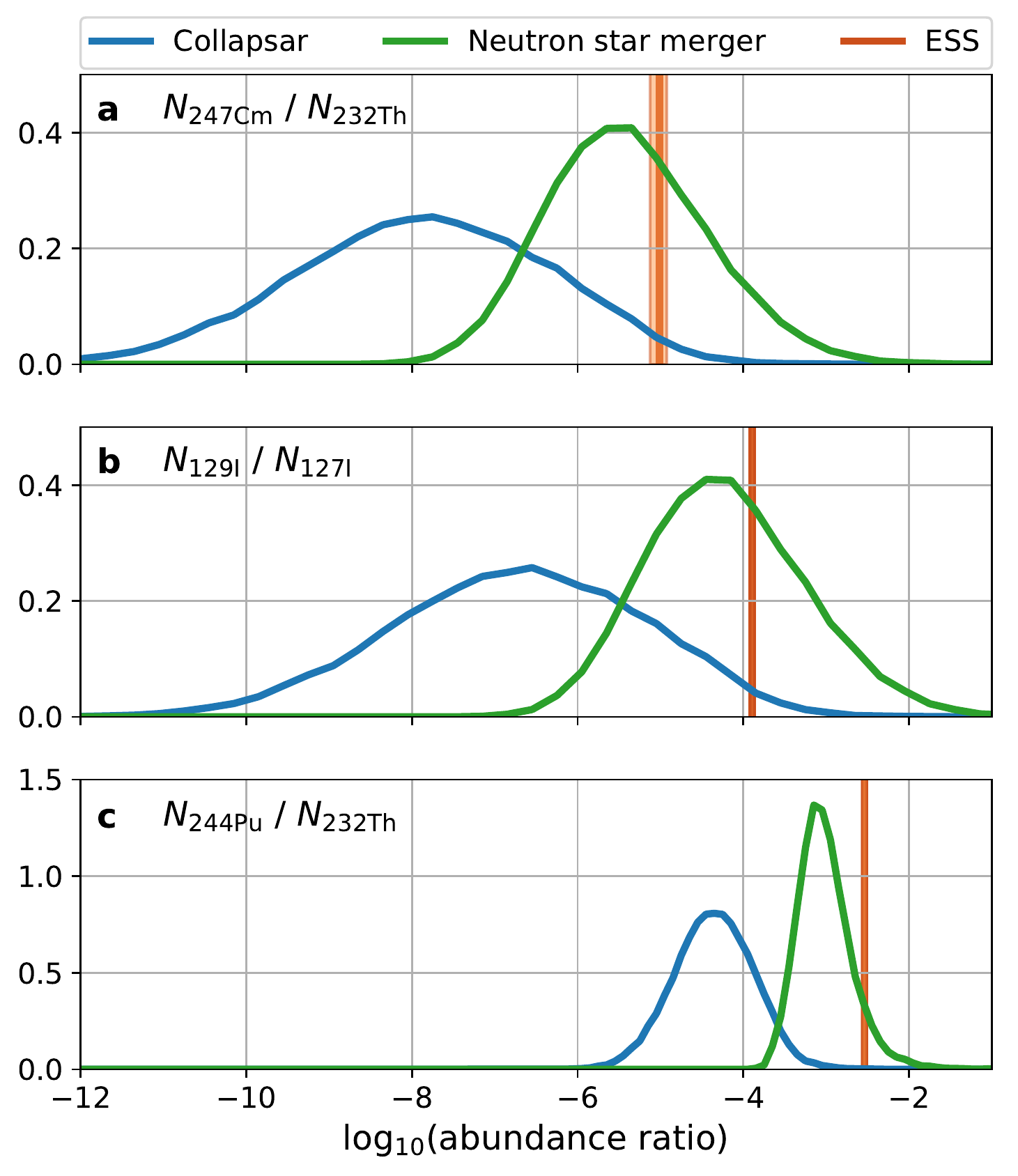}
\caption{Simulated and measured abundance ratios. Ratios $N_{\rm ^{247}Cm}/N_{\rm ^{232}Th}$ (top), $N_{\rm ^{129}I}/N_{\rm ^{127}I}$ (middle) and $N_{\rm ^{244}Pu}/N_{\rm ^{232}Th}$ (bottom) in the early Solar System are shown for simulated collapsar and neutron star merger populations, along with the measured values (see legend). The shaded area around the measured values represent $1\sigma$ uncertainties.}
\label{fig:all}
\end{figure}

\subsection{Comparison of expected abundance ratios} 

We carried out Monte Carlo simulations of neutron star mergers and collapsars in the Milky Way to calculate the expected $r$-process abundances in the early Solar System \citep{2015NatPh..11.1042H,2019Natur.569...85B}. 
We computed the abundances of Curium-247 ($^{247}$Cm; half-life $t_{1/2}=15.6$\,Myr), Iodine-129 ($^{129}$I; $t_{1/2}=15.7$\,Myr) and Plutonium-244 ($^{244}$Pu; $t_{1/2}=80.8$\,Myr) separately, normalized by the abundances of long-lived $r$-process elements such that the obtained abundance ratios could be compared to measured early Solar System values. The results, shown in Fig. \ref{fig:all}, are instructive. We see that the abundance ratio probability densities for neutron star mergers are distributed around the early Solar System values. As collapsars are more rare and, their expected contribution to the short-lived abundances is diminished, resulting in a probability distribution that is mostly much below the measured value. In addition, the collapsar rate of the Milky Way was significantly higher in its early period much before the formation of the Solar System. This injected long-lived elements into the Milky Way, further reducing the abundance ratios at the time of the Solar System's formation.

\subsection{Limit on the fractional collapsar contribution}

\begin{figure}
  \centering
   \includegraphics[width=0.49\textwidth]{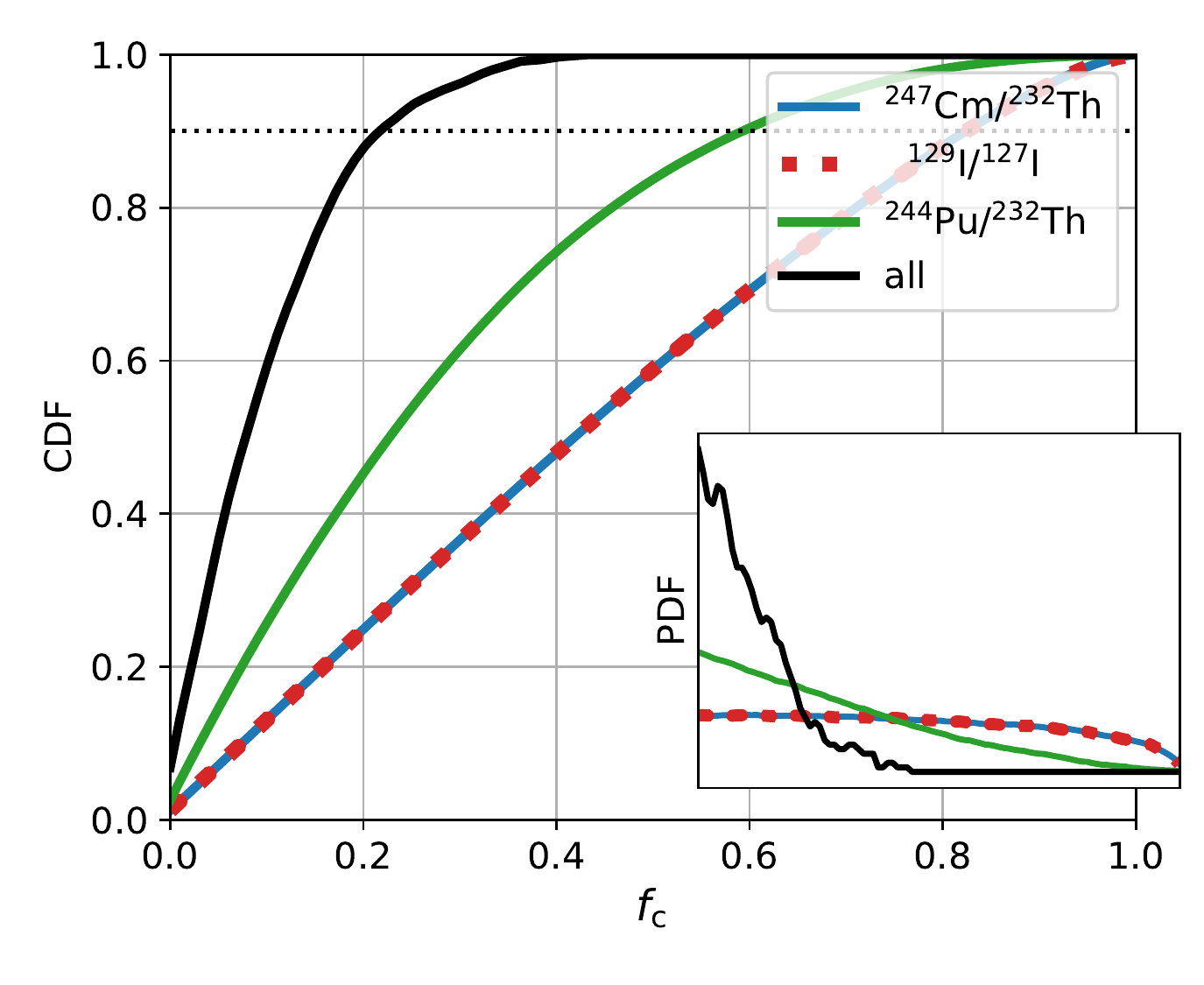}
   \caption{Cumulative probability density of fractional collapsar contribution. CDFs of the simulations agreeing with measured ratios are shown separately for $N_{\rm ^{247}Cm}/N_{\rm ^{232}Th}$, $N_{\rm ^{129}I}/N_{\rm ^{127}I}$ and $N_{\rm ^{244}Pu}/N_{\rm ^{232}Th}$ and by requiring that all three agree with observations within 30\% (see legend). The corresponding probability densities are shown in the subplot on the lower right. The horizontal dotted lines indicates 90\%.} 
   \label{fig:collapsarfractionCDF}
\end{figure}

We computed the probability density of the fraction $f_{\rm c}$ of $r$-process elements from collapsars in the Solar System. We randomly selected simulated early Solar System abundances from both the collapsar and the neutron star merger simulations, and combined them such that collapsars contribute $f_{\rm c}$ fraction of the total stable $r$-process elements. We then checked whether this combination reproduces the measured early Solar System abundance ratios to within $\pm 30\%$. The probability density of a given $f_{\rm c}$ was taken to be proportional to the fraction of simulated abundances that satisfy this criterion.

The resulting cumulative probability densities are shown in Fig. \ref{fig:collapsarfractionCDF}. We show these densities both by requiring that the abundance ratios of either a single or all elements match the measured values. From the distributions for individual mass ratios, we see that collapsar contribution is particularly constrained by $^{244}$Pu. 

\begin{figure}
  \centering
   \includegraphics[width=0.49\textwidth]{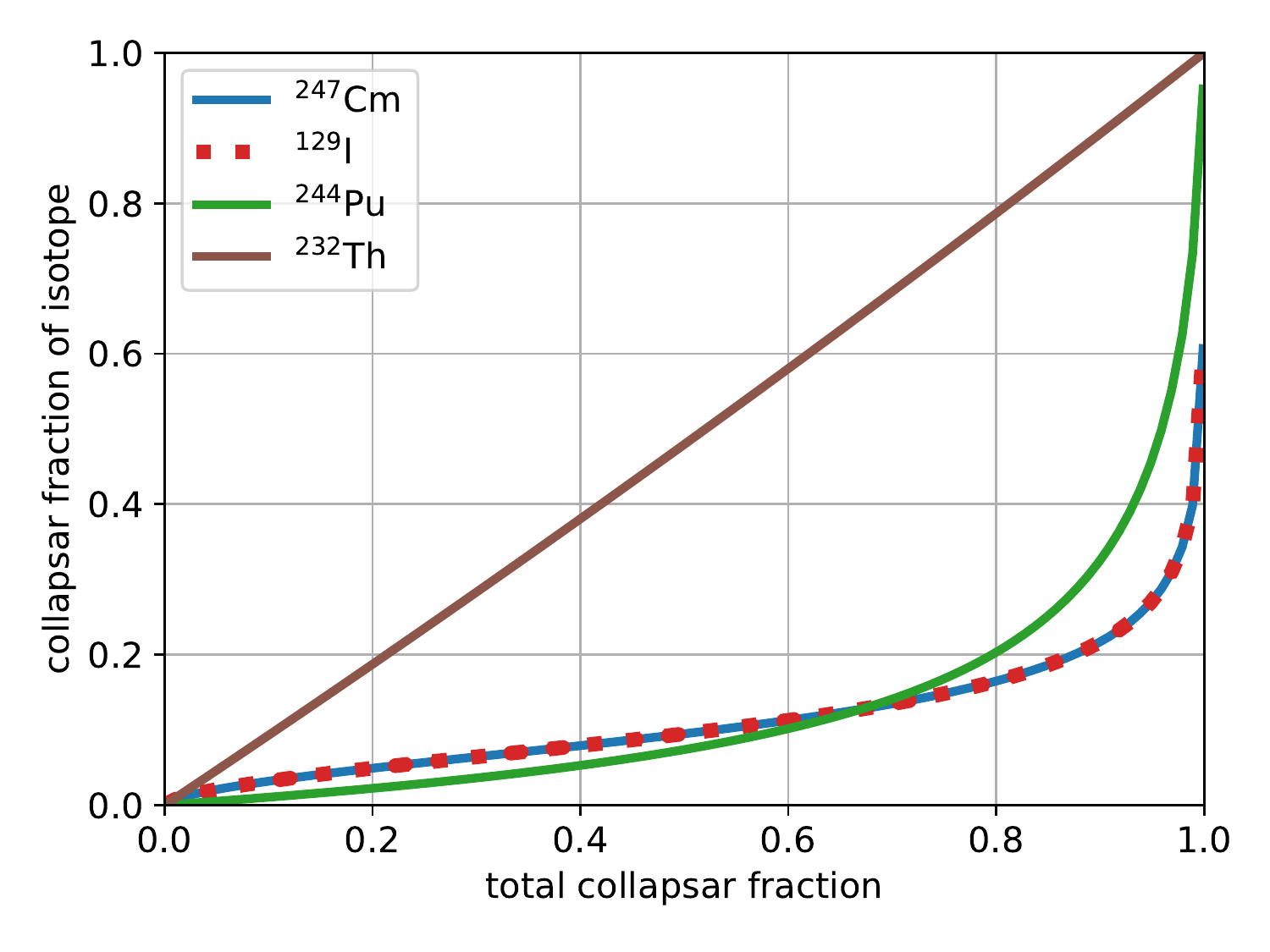}
   \caption{Fractional collapsar contribution to short lived $r$-process elements. The simulated fractions of $^{247}$Cm, $^{129}$I and $^{244}$Pu (see legend) from collapsars in the early Solar System are shown as functions of the overall fractional contribution $f_{\rm c}$ of collapsars for all $r$-process elements. } 
   \label{fig:actinideratio}
\end{figure}

To explain this, we used our simulations to compute the average fraction of different isotopes in the early Solar System that came from collapsars, as a function of $f_{\rm c}$. The obtained fractions are shown in Fig. \ref{fig:actinideratio}. We see that the fraction of long-lived isotopes is similar to $f_{\rm c}$, while we find a limited collapsar contribution to $^{244}$Pu, and essentially no contribution to $^{247}$Cm and $^{129}$I. 

By requiring all simulated abundance ratios to simultaneously match the measured early Solar System values, we see in Fig. \ref{fig:collapsarfractionCDF} that a collapsar contribution $\gtrsim 20\%$ to $r$-process elements is excluded at 90\% confidence level.

\subsection{Limits on the collapsar ejecta mass}

Using our simulations we can convert the fractional collapsar contribution $f_{\rm c}$ to the early Solar System to an estimate on the collapsar $r$-process ejecta mass. For this, we measured from our simulation the $r$-process density near the early Solar System from collapsars and neutron star mergers, in both cases normalized by their respective ejecta mass. By fixing the required collapsar fraction, we can use these abundances to find the relative ejecta masses for the two source types. We find that
For $m_{\rm ej,c}$ and $m_{\rm ej,ns}$ $r$-process ejecta masses, for collapsars and neutron star mergers, respectively, we find
\begin{equation}
m_{\rm ej,c} \approx m_{\rm ej,ns}\frac{0.6f_{\rm c}}{1-f_{\rm c}}.
\end{equation}
Taking our exclusion limit of $f_{\rm c}]\lesssim 0.2$, this means that $m_{\rm ej,c} \lesssim 0.15 m_{\rm ej,ns}$. Assuming $m_{\rm ej,ns}<0.1$\,M$_\odot$ \citep{2019arXiv190109044S}, we obtain $m_{\rm ej,c}\approx0.01$\,M$_\odot$.

\section{Conclusion} 

We found that the measured early Solar System abundances of short-lived $r$-process elements ($^{244}$Cm, $^{129}$I and  $^{244}$Pu) are typical for a neutron star merger population, while they are unlikely from a collapsar population. 

Considering contributions from both neutron star mergers and collapsars, we find that a more than 20\% collapsar contribution is excluded at 90\% confidence level for our model. This limit is due to the lower rate of collapsars compared to mergers, and their higher relative rate in the early Milky Way compared to the time of the formation of the Solar System.

Using the above limit on the fractional $r$-process contribution from collapsars and the computed deposition rate in our simulations, we exclude $r$-process ejecta masses from collapsars \Szabi{greater} than $10^{-2}$\,M$_\odot$.

The observational limits on the $r$-process ejecta mass from collapsars suggests that outflows from collapsar disks may produce less $r$-process matter than previously thought \citep{MetzgerCollapsar}. It is also possible that collapsar outflows are less neutron rich and therefore only produce lighter $r$-process elements, but not actinides. Alternatively, a currently unknown sub-population of collapsars that do not produce gamma-ray bursts would mean a higher collapsar rate \Szabi{than} considered here that would allow a higher overall fractional collapsar contribution. The results presented here will help convert future observations of kilonovae from neutron star mergers to probe the rate of collapsars, their connection to gamma-ray bursts, and the properties of accretion disks.

\begin{acknowledgments}
The authors thank Enrico Ramirez-Ruiz and Mohammad Safarzadeh for their useful feedback. The authors are grateful for the generous support of the University of Florida and Columbia University in the City of New York.
\end{acknowledgments}

\bibliography{Refs}

\end{document}